# Are Biotechnology Startups Different?

Hervé Lebret
Ecole Polytechnique Fédérale de Lausanne
Vice-Presidency for Innovation, CH-1015 Lausanne, Switzerland
Email: herve.lebret@epfl.ch – Phone: +41 21 693 70 54
Orcid : https://orcid.org/0000-0002-2420-1216

**Abstract**

In the domain of technology startups, biotechnology has often been considered as specific. Their unique technology content, the type of founders and managers they have, the amount of venture capital they raise, the time it takes them to reach an exit as well as the technology clusters they belong to are seen as such unique features. Based on extensive research from new databases, the author claims that the biotechnology startups are not as different as it might have been claimed: the amount of venture capital raised, the time to exit, their geography are indeed similar and even their equity structure to founders and managers have similarities. The differences still exist, for example the experience of the founders, the revenue and profit level at exit.

**Introduction**

Startups have been a major phenomenon of technology innovation in the last 60 years and can be dated to the beginnings of Silicon Valley in 1957. Among them, biotechnology startups have been a more recent and as important phenomenon. Their beginnings can be traced to Cetus Corporation and Genentech, both founded in the 70s, also in Silicon Valley. Today, biotechnology has become an industry in itself with some unique features. It is generally considered that the scientific content of biotechnology firms is higher than their counterparts and given the lengthy clinical trials associated with their growth, their development is slower, more costly and even riskier.

The literature on biotechnology startups is rich both from an academic as well as business-oriented standpoint. A few general books can be quoted (Shimasaki 2014, Binder 2008, Hughes 2011) and academic work is numerous even if again a few articles are worth being mentioned (Audretsch 2001; Bradford 2003; Brännback et al. 2009; Zahn and Patel 2005; Zucker et al. 1998). A recent publication by Tzabbar & Margolis (2017) exemplifies some unique features of the roles of founders.

Through research launched since 2010, the author has built two databases of information related to technology start-ups. The first one analyzes more than 5'600 Stanford-affiliated corporations and entrepreneurs (Lebret, 2017a) with a focus on fields, fundraising of the companies and background of the founders; the second one deals with more than 400 companies which had filed to go public (Lebret 2017b), with more focus on the equity structure of the companies. Both data sets include a large enough number of biotechnology startups which make it possible to revisit the specific features (if any) of these among high-tech startups.

**The features of the startups**

It seems to be common knowledge that biotechnology[1] startups would be different because of the nature of their lengthy and costly process from invention to commercialization. This would induce a longer than usual time to exit for startups as well as more fund raising from (private) investors such as venture capitalists. Some of our results are shown in Table 1 for the Stanford-related companies. The relatively surprising results are: biotechnology companies do not raise more venture capital (even if a higher percentage of firms access their capital), they do not take more time to exit (time to exit means the number of years from foundation to either an acquisition – M&A, an initial public offering - IPO or a liquidation); a much higher ratio of these goes public and fewer are acquired.

---

[1] Biotechnology (or biotech) uses living organisms to develop drugs whereas medical technologies (or medtech) deal with the development of medical instruments through electronics, mechanics and even information technologies (IT).

| Field | # | #VC-backed | % VC | VC ($M) | Time to exit | # M&As | % M&A | M&A ($M) | # IPOs | % IPO | Value at IPO ($M) | 12m. after IPO |
|---|---|---|---|---|---|---|---|---|---|---|---|---|
| Biotechnology | 258 | 171 | 66% | 44 | 6.6 | 66 | 26% | 77 | 82 | 32% | 203 | 176 |
| Medtech | 211 | 146 | 69% | 37 | 7.8 | 76 | 36% | 144 | 48 | 23% | 237 | 185 |
| IT Hardware | 879 | 485 | 55% | 49 | 7.9 | 378 | 43% | 219 | 154 | 18% | 635 | 443 |
| IT Software | 803 | 380 | 47% | 34 | 6.6 | 336 | 42% | 137 | 78 | 10% | 577 | 661 |
| Internet | 745 | 395 | 53% | 42 | 4.0 | 269 | 36% | 159 | 66 | 9% | 1'977 | 2'477 |
| Other tech. | 192 | 27 | 14% | 110 | 11.0 | 39 | 20% | 124 | 12 | 6% | 297 | 521 |
| Total[2] | 3'088 | 1'604 | 52% | 43 | 6.5 | 1'164 | 38% | 167 | 440 | 14% | 687 | 711 |

*Table 1: Features of startups by field: venture-capital (VC), years to exit, merger and acquisition (M&A) and initial public offering (IPO). Source: Stanford- affiliated corporations (Lebret 2017a).*

The explanations may be not very complex to guess as for example one unique feature of biotechnology companies is to go public on stock exchanges much earlier than others as Table 2 illustrates in terms of commercial development of the company (revenues, profits, employment). The difference in the type of exits (IPO vs. M&A) and value created at exits can be again correlated to the longer time span to profitability.

| Field | # | Sales ($M) | Income ($M) | FTEs |
|---|---|---|---|---|
| Biotech. | 122 | 10 | -17 | 76 |
| Medtech | 22 | 21 | -12 | 173 |
| HW/Comp./Tel | 68 | 92 | -7 | 432 |
| Semiconductor | 36 | 63 | -5 | 386 |
| Software | 62 | 110 | 1 | 582 |
| Internet | 106 | 336 | 3 | 1344 |
| Energy/Env. | 17 | 58 | -48 | 500 |
| Other | 6 | 169 | -15 | 391 |
| Total | 439 | 124 | -8 | 563 |

*Table 2: Additional features of startups by field: sales, incomes and employees at time of IPO (filing) Source: Lebret 2017b.*

Table 3 gives the list of the most active venture capitalists in the Stanford-related high-tech companies. Whereas biotech has been considered as a field with its own specialized investor funds, the reality is that the most active funds in biotech are in fact general funds (the ones in italics in the table).

| VC | Deals | VC investing in Biotech | Deals | VC investing in Biotech | Deals |
|---|---|---|---|---|---|
| *Kleiner Perkins Caufield & Byers* | *141* | *Kleiner Perkins Caufield & Byers* | *29* | Brentwood Venture | 5 |
| *Sequoia Capital* | *125* | *New Entreprise Associates* | *16* | Advent Venture Partners | 5 |
| *New Entreprise Associates* | *114* | Alta Partners | 14 | Arch Venture Partners | 5 |
| *Mayfield Fund* | *93* | *Venrock Associates* | *13* | Bay City Capital | 5 |
| Draper Fisher Jurvetson | 79 | Abingworth Management | 11 | Flagship Ventures | 5 |
| Accel Partners | 65 | *Mayfield Fund* | *11* | HBM Bioventures | 5 |
| *Institutional Venture Partners* | *65* | *Sequoia Capital* | *11* | Novartis Venture Funds | 5 |
| Mohr Davidow Ventures | 56 | *InterWest Partners* | *9* | Prospect Venture | 5 |
| *U.S. Venture Partners* | *56* | Asset Management Company | 8 | CDIB Bioscience | 4 |
| Menlo Ventures | 54 | *Institutional Venture Partners* | *8* | Charter Life Sciences | 4 |
| Sutter Hill Ventures | 53 | *Morgenthaler Ventures* | *8* | Montreux Equity | 4 |
| *Venrock Associates* | *50* | MPM Capital | 8 | Skyline Ventures | 4 |
| *InterWest Partners* | *48* | Domain Partners | 7 | GSK | 3 |
| Greylock Partners | 48 | Sofinova Partners | 7 | Longitude Capital | 3 |
| Benchmark Capital | 47 | Versant Ventures | 7 | OrbiMed Advisors | 3 |
| *Morgenthaler Ventures* | *39* | Aberdare Ventures | 6 | Pfizer | 3 |
| Norwest Venture Partners | 37 | *Oak Investment Partners* | *6* | Polaris | 3 |
| Bessemer Venture Partners | 36 | Oxford Bioscience Partners | 6 | Rho Management | 3 |
| *Oak Investment Partners* | *35* | | | SR One | 3 |
| *Alta Partners* | *33* | | | *U.S. Venture Partners* | *3* |

*Table 3: Most active venture capital funds overall and in biotechnology. Source: Lebret 2017a.*

There are however specialized funds in biotech such as Abingworth Management, Asset Management Company, MPM Capital, Domain Partners, Sofinova Partners, Versant Ventures, Aberdare Ventures or Oxford Bioscience Partners among the most active ones (but this is not different from other fields such as IT which also has its dedicated funds such as Andresseen Horowitz, the Founders Fund or Benchmark Capital). Kleiner Perkins

---

[2] The 5'800 corporations affiliated to Stanford University include companies in the fields of finance, law, consulting, and are not considered as high-tech start-ups.

Caufield & Byers is the most active fund and interestingly enough was a fund which invested in both Cetus and Genentech, the "first" biotechnology startups ever…

**The features of the founders and management teams**

The two databases provide also interesting information about the founders, managers and equity structure of the startups. Table 4 shows the relatively higher percentage of professors, PhDs and lower percentage of MBAs among biotechnology founders whereas Table 5 illustrates the higher age of founders, their smaller equity ownership at IPO as well as the higher ownership of investors and new shareholders at IPO.

| Field | Founding professors | | Founding PhDs | | Founding MBAs | | Licenses from Stanford | |
|---|---|---|---|---|---|---|---|---|
| Biotechnology | 70 | 27% | 62 | 24% | 45 | 17% | 83 | 32% |
| Medtech | 44 | 21% | 21 | 10% | 38 | 18% | 42 | 20% |
| IT Hardware | 83 | 9% | 304 | 35% | 115 | 13% | 50 | 6% |
| IT Software | 48 | 6% | 174 | 22% | 191 | 24% | 28 | 3% |
| Internet | 19 | 3% | 76 | 10% | 276 | 37% | 5 | 1% |
| Other tech | 6 | 3% | 32 | 17% | 49 | 26% | 6 | 3% |
| Total | 270 | 9% | 669 | 22% | 714 | 23% | 214 | 7% |

*Table 4: Involvement of Stanford professors, PhDs and MBAs. Source: Lebret 2017a.*

| Field | Founders' Age | Ownership | | | | Non founding CEOs | |
|---|---|---|---|---|---|---|---|
| | | Founders | Employees | Investors | IPO | # | % |
| Biotech. | 45 | 7% | 15% | 57% | 22% | 52 | 43% |
| Medtech | 42 | 8% | 18% | 52% | 22% | 8 | 36% |
| HW/Comp./Tel | 37 | 13% | 28% | 45% | 16% | 26 | 38% |
| Semiconductor | 38 | 12% | 26% | 44% | 18% | 15 | 42% |
| Internet | 33 | 17% | 22% | 48% | 14% | 33 | 31% |
| Software | 33 | 17% | 27% | 41% | 17% | 16 | 26% |
| Energy/Env. | 37 | 8% | 20% | 55% | 18% | 7 | 41% |
| Other | 39 | 10% | 21% | 52% | 20% | 3 | 50% |
| Total | 38 | 13% | 22% | 49% | 18% | 160 | 36% |

*Table 5: Features about founders, non-founding CEOS and other shareholders. Source: Lebret 2017b.*

One can link the higher status of founders (professors, PhDs) in biotechnology with their age, when founding companies and this seems to be also an illustration of the higher scientific content of the startups or longer experience of the founders, which can exist after many years of research only. As a non-surprising and correlated note, the non-founding CEOs are relatively more frequent. Professors seldom leave their academic position and do not have full-time, not even executive positions in the startups, but have often advisory roles.

| Region | Biotech | Medtech | Software | Internet | HW/Comp./Tel | Semicon. | Energy/Env. | Other | Total |
|---|---|---|---|---|---|---|---|---|---|
| Silicon Valley | 15 | 4 | 13 | 26 | 20 | 7 | 6 | 2 | 93 |
| Boston Area | 22 | 2 | 5 | 5 | 3 | | 3 | 1 | 41 |
| Southern California | 17 | 5 | | 1 | | 2 | 1 | | 26 |
| West Coast | 3 | | 2 | 3 | 2 | 1 | 1 | | 12 |
| East Coast | 20 | | 3 | 5 | 1 | | 2 | | 31 |
| Midwest | 2 | | 2 | 3 | 1 | | 2 | 1 | 11 |
| Switzerland | 7 | 2 | 1 | | 1 | | | | 11 |
| United Kingdom | | | 2 | 3 | | | | | 5 |
| France | 4 | 3 | 1 | 3 | 5 | 2 | 1 | | 19 |
| Other EU | 1 | | 3 | 3 | 2 | | | | 9 |
| China | 1 | | 1 | 8 | | | | | 10 |
| Israel | | | 1 | | 1 | | 1 | | 3 |
| Rest of the World | 1 | | 2 | 2 | | | | | 5 |
| Total | 93 | 16 | 36 | 62 | 36 | 12 | 17 | 4 | 276 |

*Table 6: Geographic information about 276 firms which had filed for an IPO since 2007. Source: Lebret 2017b.*

The two main technology clusters are well-known to be Silicon Valley and the Boston area. Indeed Boston is argued to be the biotech cluster. The numbers in Table 6 slightly mitigate that point of view as California as a whole has more start-ups than Boston but it remains a fact that Boston has a higher proportion of biotech startups than Silicon Valley, but not more than Southern California though.

**Features about licensing**

Licensing of technologies for high-tech companies is an important topic. Many not to say most high-tech start-ups are dependent at foundation upon intellectual property (IP) such as patent applications. The last column of

Table 4 shows that biotechnology companies are even more dependent than others upon IP licensing from Stanford University. The database of companies having filed to go public shows similar results. Out of the 430 filings, 38 companies had indicated licensing technologies from a university, out of which 24 were biotechnology companies (63% of the total even if this last figure may not be statistically relevant).

The license terms are generally a combination of equity in the startup and royalties on sales. It is interesting to notice that the average equity stake of universities in biotech startups is 10% at creation and 5% after series A (of an average size of $6M), numbers which are typical of academic licenses to startups in any technology field (Lebret 2017b). The main difference in academic licenses with biotechnology startups has not to do with the equity component but with the royalty one. Royalties typically range from 2% to 4% of revenues based on licensed IP in biotech and this even replaces the equity component in some cases. Royalties are (known to be) much less accepted in other fields. This might have to be related to the revenue amount at time of exits which are much lower in biotech than in other fields so that managers and investors in biotech might be less sensitive to that last feature.

**As a conclusion**

Even if the author tried to provide mainly facts and figures from two specific databases, it is interesting to try and interpret this data. The higher scientific content of biotech startups seems to be confirmed by the nature and status of their founders. Their lengthy development is not confirmed during their phases as private firms but only through their situation at IPO: biotech firms go public with less commercial validation, and therefore before the end of their clinical trials. The IPO is just another step in their funding process and not a final validation of their commercial prospects.

Biotech are embedded in the same technology clusters as the other high-tech startups and they are often funded by the same venture capitalists. In any case, startups have developed historically in parallel with venture capital, and this industry has indeed shaped the high-tech corporations through their own model. With a ten-year life span VC funds must experience exits of their portfolio company during the fund life. They have also shaped the management and equity structure of these companies to maximize their likeliness of success. These facts may be illustrations of why biotech startups are not very different from other technology startups, despite their different initial content.